# CRC-ICM: Colorectal Cancer Immune Cell Markers Pattern Dataset

Zahra Mokhtari[1], Elham Amjadi[2], Hamidreza Bolhasani[3], Zahra Faghih[4], AmirReza Dehghanian[5], Marzieh Rezaei[1]*

1. Department of Immunology, School of Medicine, Isfahan University of Medical Sciences, Isfahan, Iran.
2. Poursina Hakim Digestive Diseases Research Center, Isfahan University of Medical Sciences, Isfahan, Iran
3. Department of Computer Engineering, Science and Research Branch, Islamic Azad University, Tehran, Iran.
4. Institute for Cancer Research (ICR), School of Medicine, Shiraz University of Medical Sciences, Shiraz, Iran
5. Department of Pathology, School of Medicine, Shiraz University of Medical Science, Shiraz, Iran

**\* Corresponding Author:** Dr. Marzieh Rezaei, E-mail: ma.rezaei@med.mui.ac.ir

**Abstract**

Colorectal cancer (CRC) is the second most common cause of cancer death in the world, and can be identified by the location of the primary tumor in the large intestine: right and left colon, and rectum. Based on the location, CRC shows differences in chromosomal and molecular characteristics, microbiomes, incidence, pathogenesis, and outcome. It has been shown that tumors on the left and right sides also have different immune landscapes, so the prognosis may be different based on primary tumor locations. It is widely accepted that the immune components of the tumor microenvironment (TME) play a critical role in tumor development. Accordingly, analysis of the interactive relationships between tumor cells and the immune system components in the TME have received more attention, while routine grading system such as TNM classification does not have the power to predict the clinical outcome of the disease. Therefore, the identification of immune-related markers in TME might help to predict the prognosis and clinical outcome of the disease. One of the critical regulatory molecules in the TME is immune checkpoints that as the gatekeepers of immune responses regulate the infiltrated immune cell functions. Inhibitory immune checkpoints such as PD-1, Tim3, and LAG3 as the main mechanism of immune suppression in TME overexpressed and result in further development of the tumor. Therefore, this research aimed at providing a well-organized histopathological microscopy image dataset for best-scoring CRC. For this purpose, the images have been taken from colon tissues of patients with CRC stained with specific antibodies for CD3, CD8, CD45RO, PD-1, LAG-3, and Tim3 and separately determined in both invasive margin and center of tumor for each marker. Moreover, the present dataset contains comprehensive information on the clinicopathological characteristics of the patients, focusing on the location of the tumor, separating the right and left. The name of this dataset is CRC-ICM and contains 1,756 images related to 136 patients. The dataset includes an excel file describing patient's demographic, anatomical features, histopathological grade and some more technical details. The initial version of CRC-ICM is published on the Elsevier Mendeley dataset portal, and the latest version is accessible via: https://databiox.com

# Introduction

Colorectal cancer (CRC), containing tumors arising from the colon and/or rectum, is a major public health concern as it is the third most commonly diagnosed and second deadliest cancer worldwide. Approximately 9.4% of cancer-related deaths in 2020 were due to CRC [1]. Surgery is the first line of treatment for patients with localized CRC in stages I-III while adjuvant chemotherapy is prescribed to eliminate micro-metastases and prevent tumor recurrence after surgery and lumpectomy [2, 3].

Now, AJCC/UICC TNM staging system is the gold-standard method for classification of CRC cancer. This method relies on tumor features including, the extent of the primary tumor (T), the regional lymph nodes involvement (N), and the distant metastasis (M) [4]. Despite its global acceptance, importance, and power, TNM staging has several drawbacks as clinical outcomes and prognoses may vary among patients with the same stage [4, 5].

As a result of genomic instability, each tumor has a unique molecular profile, so explaining the vastly different prognosis and response to the treatment of individual patients. Understanding of this molecular heterogeneity is essential for the optimal risk stratification of patients and the implementation of current and future therapeutic strategies [6].

It is now widely accepted that the immune component of the tumor microenvironment (TME) plays an important role in tumor development [7]. Therefore, analysis of the interactions between tumor cells and components of the immune system in TME has received more attention [8]. Many studies have shown that high densities of various T cell subpopulations, such as CD3+ T cells, CD8+ T cells, and CD45RO+ memory T cells, in tumor tissue are associated with longer overall survival (OS) and disease-free survival (DFS) [8]. Accordingly, Galon designed and proved a method as immunoscore, provides a measure based on the density of two CD3+/CD8+ lymphocyte populations in the center and the invasive margin of the tumor. This method can be placed next to the TNM system in order to better classify patients with CRC [9, 10].

Immune checkpoints (ICPs) refer to paired receptor-ligand molecules that as gatekeepers of the immune response could exert activatory, inhibitory or dual effects on immune system. In TME, infiltrating immune cells exhibit a widespread dysfunction characterized by inhibitory signals [11, 12]. The roles of these immune cells in tumor growth and progression are diverse and closely related to the molecules and ligands they express [13]. Among them, upregulation of inhibitory immune checkpoints (iICPs) are hallmarks of the tumor ecosystem, leading to immune cell dysfunction [14]. Thus, in addition to the extent of immune cell infiltration, the function of these cells should be assessed to better understand the state of the immune system in the TME.

Some iICPs are include programmed cell death receptor 1 (PD-1), T cell immunoglobulin domain and mucin domain-containing 3 (Tim3), lymphocyte activation 3 (LAG3), cytotoxic T lymphocyte-associated protein-4 (CTLA4), a cellular immune receptor with T Ig and ITIM domains (TIGIT) and B and T lymphocyte attenuator (BTLA). Their expression in malignant tumors is often markedly increased and associated with poor prognosis [14]. Considering the importance of these molecules, inhibition of these iICPs with specific antibodies is one of the main strategies of immunotherapy. Currently, three antibodies Pembrolizumab, Nivolumab (PD-1 inhibitors), and Ipilimumab (CTLA-4 inhibitor which is used along with Nivolumab) in metastatic CRC patients who have high microsatellite instability (MSI) are FDA-approved [15].

Despite the importance of the immune system in tumor progression, many aspects of the immune system in cancer are still unknown. Much research on the immune system and immune-based therapies need to be conducted to achieve better clinical outcomes and personalized treatment for patients. Additionally, the data collected and organized from the study will be made available to other researchers in the form of datasets that can be effectively used in the design and analysis of future studies.

In the present study, we introduced the dataset containing immunohistochemical data of immune markers including CD3, CD8, CD45RO, PD-1, Tim3, and LAG3. The pictures were taken from both the invasive margin (IM) and center of tumor (CT) at 200x magnification. According to our knowledge, this is the first dataset containing images of PD-1, Tim3, and LAG3 expression from the IM and CT separately, besides the expression of immunoscore markers in patients with CRC.

**Method & Result**

Colorectal cancer patients who underwent surgery at Al-Zahra Hospital (Isfahan University of Medical Sciences, Iran) between 2013 and 2016 were selected. Patients who didn't receive preoperative chemotherapy, hadn't a history of other cancers or autoimmune diseases, and had appropriate tissue were included. An experienced pathologist examined the Hematoxylin and Eosin (H&E) slides to confirm the patient's pathological data and select the optimal tissue block containing both the center and invasive margin of the tumor.

Immunohistochemistry (IHC) staining for CD3, CD8, CD45RO, PD-1, Tim3, and LAG3 was performed on 4 µm sections of paraffin-embedded tissue (more details in [16, 17]). Two experienced pathologists then determined the invasive margin from the tumor center and tacked images of both areas (in the IHC procedure, some tissues miss the CT or IM).

Finally, we created a database named CRC-ICM contains 1,756 images related to CRC patients. The dataset includes an Excel file describing the patient's demographic, anatomical features, histopathological grade and some more technical details. overall, the study included 136 CRC patients (male= 83, female= 53) with a mean age (±SD) at diagnosis of $62.35 \pm 14.10$ years (range 19–92 years). Based on tumor location, patients were divided into two main categories: right (cecum, ascending, liver flexure, transverse colon) and left (splenic flexure, descending sigmoid, rectosigmoid, rectum). Most patients were in pathological stage II (n=46, 33.8%) and stage III (n=40, 29.4%). 90 patients (66.2%) had no lymph node involvement and 124 patients (91.2%) had no distant metastases (M0) at the time of surgery. Complete patient demographics are summarized in Table 1.

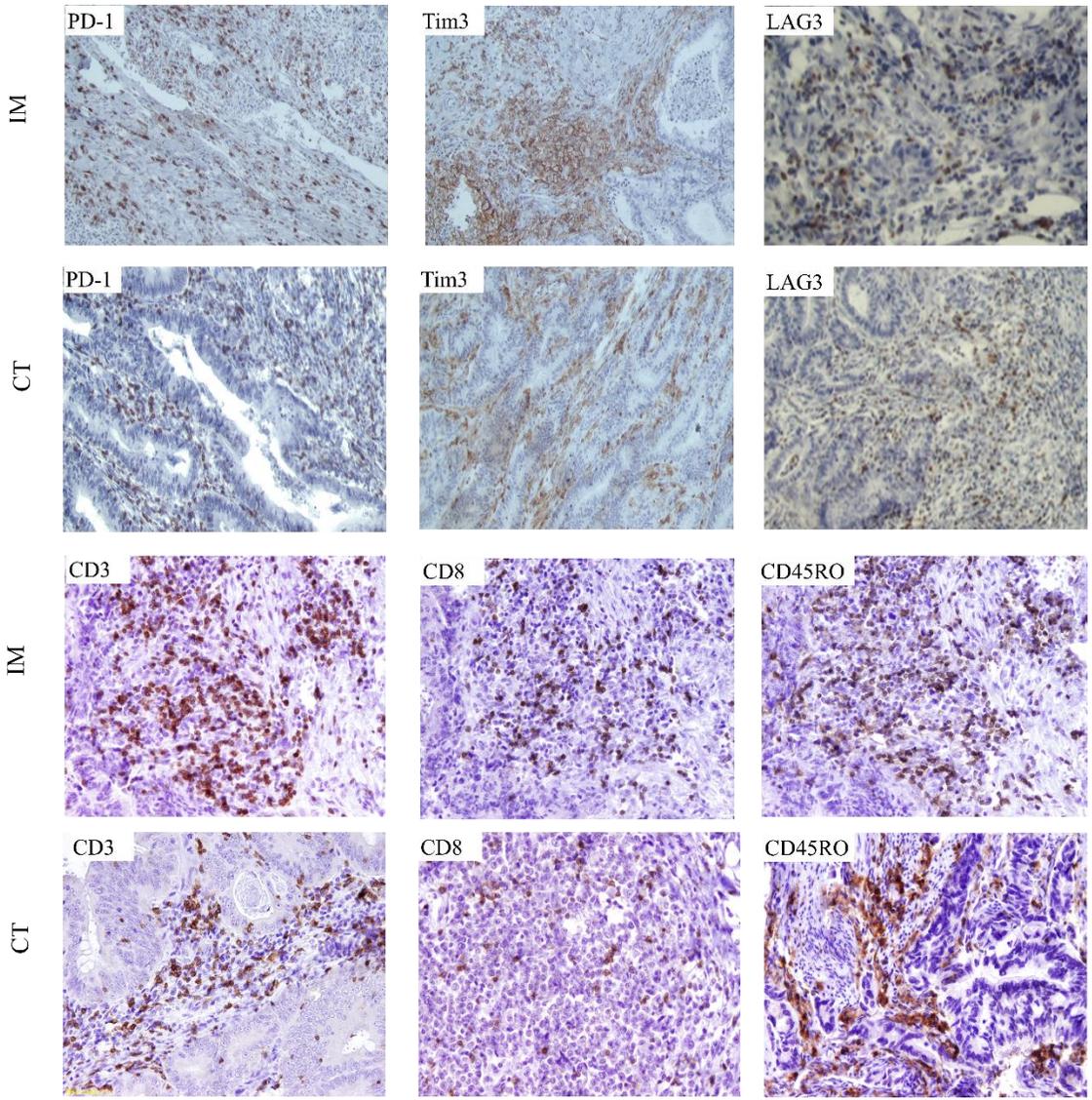

**Figure 1.** IHC staining of PD-1, Tim-3, LAG3, CD3, CD8, and CD45RO expressions in CRC (200×).
**CT:** Center of tumor  **IM:** Invasive margin

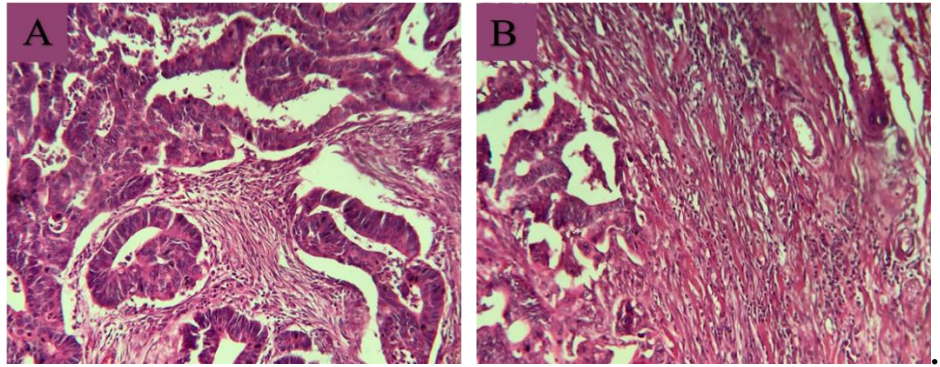

**Figure 2.** H & E staining of CRC tissues in center of the tumor (CT) **(A)** and invasive margin (IM) **(B)**.

The data of this study includes images of slides stained with CD3, CD8, CD45RO, PD-1, Tim3, LAG3, and H&E from 96 patients on both IM and CT are available in this dataset. For 17 patients, only images related to CD3, CD8, and CD45Ro are available. Eighteen patients only have images associated with PD-1, Tim3, LAG3, and H&E slides. The number of images for each marker and tumor region is listed in Table 2, and some technical specifications of this dataset are presented in Table 3. Some image samples of IHC staining of PD-1, Tim-3, LAG3, CD3, CD8, and CD45RO expressions in CRC with 20x magnification level and also Hematoxylin and Eosin staining of CRC tissues in center of the tumor (CT) and invasive margin (IM) are presented in Figure 1 and 2, respectively. The initial version of CRC-ICM is published on the Elsevier Mendeley dataset portal: https://data.mendeley.com/datasets/h3fhg9zr47/1. And also, the latest version is accessible via DataBioX website: https://databiox.com.

**Table 1.** Demographics characteristics of the patients in CRC-ICM dataset

| Parameters | No. of Cases (%) | Parameters | No. of Cases (%) |
|---|---|---|---|
| **Total** | 136 | **TNM stage** | |
| **Sex** | | 0/I | 38 (28.0) |
| Male | 83 (61.0) | II | 46 (33.8) |
| Female | 53 (39.0) | III | 40 (29.4) |
| **Age** | | IV | 12 (8.8) |
| <63 | 65 (47.8) | **Lymphovascular Invasion (LVI)** | |
| ≥63 | 71 (52.2) | Absent | 79 (58.1) |
| **Tumor side** | | Present | 57 (41.9) |
| Right | 56 (41.2) | **Perineural invasion** | |
| Left | 76 (55.9) | Absent | 111 (81.6) |
| Unknown | 4 (2.9) | Present | 25 (18.4) |
| **Tumor size** | | **Metastasis** | |

| <5 | 56 (41.2) | Absent | 89 (65.4) |
|---|---|---|---|
| ≥5 | 78 (57.4) | Present | 34 (25.0) |
| Unknown | 2 (1.5) | Unknown | 13 (9.6) |
| **Differentiation grade** | | **Recurrence** | |
| Low grade | 76 (55.9) | Absent | 97 (71.3) |
| Moderate grade | 53 (39.0) | Present | 29 (21.3) |
| High grade | 7 (5.1) | Unknown | 10 (7.4) |
| **T stage** | | **Tumor budding** | |
| T1/Tis | 4 (3.0) | Low | 89 (65.4) |
| T2 | 43 (31.6) | High | 47 (34.6) |
| T3 | 72 (52.9) | **Tertiary lymphoid structure (TLS)** | |
| T4 | 17 (12.5) | Absent | 104 (76.5) |
| **Lymph node involvement** | | Present | 32 (23.5) |
| Absent | 90 (66.2) | **Survival** | |
| Present | 46 (33.8) | Alive | 79 (58.1) |
| **M stage** | | Dead | 57 (41.9) |
| M0 | 124 (91.2) | | |
| M1 | 12 (8.8) | | |

**Table 2.** Number of images for each marker and tumor region

| | Number of Images | |
|---|---|---|
| | CT | IM |
| **CD3** | 113 | 113 |
| **CD8** | 113 | 113 |
| **CD45RO** | 113 | 113 |
| **PD-1** | 114 | 114 |
| **Tim3** | 114 | 114 |
| **LAG3** | 114 | 114 |
| **H&E** | 114 | 114 |
| **Overall** | 1756 | |

**Table 3.** CRC-ICM dataset technical specifications.

| Dataset Item | Value |
| --- | --- |
| Number of Images | 1756 |
| Image Format | RGB |
| Image File Type | JPG |
| Image Resolution | 72 ~ 96 dpi |
| Size (Pixels) | 4140 * 3096<br>2070 * 1548<br>1280 * 960 |

**Discussion**

CRC is a heterogeneous disease with high prevalence and mortality. That is why improving the classification and treatment of patients and moving towards personalized medicine is so important. Due to the critical role of the immune system in tumorigenesis, understanding the immune aspects of tumors may be key to achieving these goals. In recent years, many studies have been conducted on the relationship between immune markers and cancer treatment and prognosis including CRC. However, a well-organized dataset is a mandatory requirement for training machine learning algorithms. In the CRC-ICM dataset, images of six key immune markers were collected, indicating the degree of penetration and functional status of immune cells in TME.

Moreover, protein expression and functional activity of various signaling pathways have been described as differences between CT and IM of CRC. Thus, separate CT and IM images were obtained for a more comprehensive landscape of the immune system status. Intra-tumor heterogeneity is well-characterized for various features of the tumors, and many studies have been conducted examining the difference between the IM and the CT of the tumor [18, 19]. Several prognostic and biologically important markers such as MACC1, E-cadherin, and ZEB2 are differentially expressed between the IM and the CT [20-22]. Therefore, different regions of the tumor, including the IM and the CT, should be evaluated for a better and more complete examination of the tumor.

In addition to comprehensive images, these type dataset provides complete clinical and pathological information about patients, allowing researchers to assess patients in different subgroups. For example, right and left-sided CRC differ in molecular signaling pathways, prognosis, and response to treatment [23]. Therefore, grouping patients accordingly can be very important.

In order to promote related research, we release CRC-ICM-v1 dataset contains clinical data and unaligned histopathological, and immune cell markers images of 136 CRC patients. This dataset opens the door for future pathology studies on CRC and the immune cell interaction in TME.

**Table 4.** Abbreviations and Definitions.

| Abbreviation | Definition | Abbreviation | Definition |
|---|---|---|---|
| **CRC** | Colorectal Cancer | **Tim3** | T cell Immunoglobulin Domain and Mucin Domain-Containing 3 |
| **TME** | Tumor Microenvironment | **LAG3** | Lymphocyte Activation 3 |
| **OS** | Overall Survival | **CTLA4** | Cytotoxic T Lymphocyte-Associated Protein-4 |
| **DFS** | Disease-Free Survival | **H & E** | Hematoxylin and Eosin |
| **ICPs** | Immune Checkpoints | **IHC** | Immunohistochemistry |
| **iICPs** | Inhibitory Immune Checkpoints | **MSI** | Microsatellite Instability |
| **PD-1** | Programmed Cell Death Receptor 1 | **HMGB1** | High Mobility Group Protein B1 |
| **PtdSer** | Phosphatidylserine | **NK-Cells** | Natural Killer Cells |
| **TILs** | Tumor-Infiltrating Lymphocytes | **CT** | Center of the Tumor |
| **IM** | Invasive Margin | | |

DECLARATIONS

AVAILABILITY OF DATA AND MATERIALS

Available.

FUNDING

Not Applicable.

ACKNOWLEDGEMENTS

Not Applicable.

CONFLICT OF INTEREST

o   All authors have participated in (a) conception and design, or analysis and interpretation of the data; (b) drafting the article or revising it critically for important intellectual content; and (c) approval of the final version.

o   This manuscript has not been submitted to, nor is under review at, another journal or other publishing venue.

o   The authors have no affiliation with any organization with a direct or indirect financial interest in the subject matter discussed in the manuscript

o   The following authors have affiliations with organizations with direct or indirect financial interest in the subject matter discussed in the manuscript: